# Comment

**Arnold Zellner**

The authors are to be congratulated for their deep appreciation of Jeffreys's famous book, *Theory of Probability*, and their very impressive, knowledgeable consideration of its contents, chapter by chapter. Many will benefit from their analyses of topics in Jeffreys's book. As they state in their abstract, "Our major aim here is to help modern readers in navigating this difficult text and in concentrating on passages that are still relevant today." From what follows, it might have been more accurate to use the phrase, "modern well-informed Bayesian statisticians" rather than "modern readers" since the authors' discussions assume a rather advanced knowledge of modern Bayesian statistics. Readers who are "just" physicists, chemists, philosophers of science, economists, etc., may have great difficulty in understanding the authors' guide to Jeffreys's book. This is unfortunate since the book provides methods and philosophical principles relevant for all the sciences. Perhaps in the future, additional reviews of Jeffreys's book will be prepared that are understandable to a broader range of readers, as was done in having scientists and scholars from many fields discuss at length Jeffreys's and others' thoughts on simplicity and complexity at a conference and reported in Zellner, Kuezenkamp and McAleer (2001).

Another point that affects the authors' discussion is their apparent misinterpretation of the title of Jeffreys's book. They write, "The title itself is misleading in that there is no exposition of the mathematical bases of probability theory in the sense of Billingsley (1986) and Feller (1997)." In this regard, years ago Lord Rutherford, a famous physical scientist, said that if you need statistics to analyze your data, you better redesign your experiment, and as a result the word "statistics" was not highly regarded in the physical sciences and the term "probability theory" was employed by Jeffreys, Jaynes (2003) and many other physical scientists to include applied and theoretical statistics, mathematical methods, including elements of formal probability theory and philosophical aspects of science. With their narrow interpretation of Jeffreys's title, the authors found many discussions in the book to be "irrelevant," whereas Jeffreys considered them to be of fundamental importance and did not want to have his book limited to just mathematical topics, as in his and his wife's very famous book, *Mathematical Methods of Physics*. And indeed, Good [(1980), page 32] wrote, "In summary, Jeffreys's pioneering work on neo-Bayesian methods... was stimulated by his interest in philosophy, mathematics, and physics, and has had a large permanent influence on statistical logic and techniques. In my review Good (1962) I said that Jeffreys's book on probability "is of greater importance for the philosophy of science, and obviously of greater immediate practical importance, than nearly all the books on probability written by professional philosophers lumped together." I believe this is still true, though more professional philosophers have woken up."

With respect to the discussion of Chapter 1, readers will wonder what the authors mean by terms like "subjective," "objective," "objective priors" and "genuine prior information." Contrary to what the authors state, Jeffreys did adjust his "objective priors" (1) to get a "reasonable" amount of invariance, (2) to get "reasonable" results in the Laplace rule of succession, binomial problem and (3) to correct for "selection results" in testing many alternative models with large sets of data. Thus he was not always an "objective" Bayesian but rather a very thoughtful Bayesian who recognized needs for better procedures for certain problems and provided them in many cases. Perhaps he should be called a "pragmatic" Bayesian.

Most important in Chapter 1 is Jeffreys's axiom system for learning from data and experience that is applicable to research in all fields of science. He considered deduction and induction at great length in a most interesting productive manner and the authors provide interesting and useful comments. However, the authors' introduction of decision theoretic considerations as a solution in discussion of point 1 fails to recognize that the decision theoretic solution based on limited data, though "optimal" may not be very good because of the limited data employed. Good science requires testing models' explanatory and predictive performance using much data in order to ascertain the validity of a particular theory, say Einstein's theory and along the way in testing many variants of the original model will probably be considered. And finally, one has to specify a loss or utility function... whose loss function? Errors in formulating loss or utility functions can vitally affect the quality of "optimal" decisions, as is well known. And to suggest that the debate about model choice was not present in Jeffreys's time overlooks the well-known debates that raged about Newton's "laws" versus Einstein's "laws" and the adequacy of quantum theory, etc., during the early 20th century and beyond, about which Jeffreys was fully aware. Further, the authors' statement in point 8 about grounding *Theory of Probability* within mathematics fails to note that Jeffreys recognized that there is controversy about the foundations of mathematics. Still he pragmatically adopted point 8.

*Arnold Zellner, H.G.B. Alexander Dist. Service Prof. Emeritus, Booth Business School, U. of Chicago, 5807 South Woodlawn Avenue, Chicago, Illinois 60637, USA e-mail: Arnold.Zellner@chicagobooth.edu; URL: http://faculty.chicagobooth.edu/arnold.zellner/more/index.htm.*

Most important in Chapter 1 are Jeffreys's comments on his dissatisfaction with the standard proof of the product rule of probability that is used to derive Bayes' theorem that led him to introduce the product rule of probability as an axiom, rather than a theorem in his system, as the authors note in their discussion of the derivation of Bayes' theorem. Jeffreys noted that the assumption that the elements of the sets A, B and the intersection of A and B are equally likely to be drawn, all having a probability equal to $1/n$, where $n$ is the total number of elements, will not be satisfied in many cases. After stating that he was unable to prove the product rule without this assumption, he pragmatically introduced the result as Axiom 7 on page 25. Since many, including myself, worried about this basic point, I was happy to discover that the proof of the product rule could be generalized by going to a hierarchical model with the probabilities for elements of the sets assumed to have properties that produced the usual product rule of probability; see Zellner (2007). Further, earlier in my concern about valid proofs or derivations of Bayes' theorem, I approached the problem as an engineer might by considering the informational inputs, namely the information in a prior density and in a likelihood function, and the output information, the information in a posterior density for the parameters and a marginal density for the observations. On using Shannon's measure of information, it is possible to form an expression, output information minus input information and to minimize it with respect to the choice of the form of the output or posterior density for the parameters. The solution is to take the posterior density equal to the prior density times the likelihood function divided by the marginal density of the observations, which is precisely the result yielded by Bayes' theorem. Also, when this solution is employed, it is the case that the output information equals the input information and thus the procedure is 100% efficient. See Zellner (1988) for the detailed results and commentary on them by E. T. Jaynes, B. M. Hill, S. Kullback and J. Bernardo, all reprinted in Zellner (1997b) along with solutions to variants of the above problem. For example, in some problems we may not have an input prior but just an input likelihood function. Then the solution to the minimization problem is to take the posterior density proportional to likelihood function, a 100% efficient solution that happens to be exactly the fiducial inference procedure suggested by R. A. Fisher who, as Jeffreys and others pointed out, did not have a theoretical justification for it. Also, other optimal information processing results are presented that take account of the varying quality of input information, temporal relations of the inputs from one period to the output of the next period, etc. In effect, we now have a number of optimal learning models, not just one, Bayes' theorem, to use in learning from data and experience. Given that Jeffreys was deeply concerned about how to justify Bayes' theorem and how to learn effectively from data and experience, I hope that he likes these results that flowed from his concern about the validity and applicability of proofs of Bayes' theorem.

With respect to the authors' comments on prior densities, in particular non-informative priors, they very thoughtfully review Jeffreys's innovative procedure for producing non-informative priors with many critical remarks regarding his use and misuse of unbounded measures. As regards a prior for the binomial parameter $p$, which can take on values in the closed interval zero to 1, the authors consider

Laplace's uniform prior, Haldane's prior and the Jeffreys's prior followed by a thoughtful discussion of the famous Laplace Rule of Succession for analysis of which Jeffreys suggests putting lumps of probability on the values zero and one and spreading out the remaining probability mass uniformly, zero to one in order to get "reasonable" results for Laplace's problem: given $n$ independent dichotomous, binomial trials and observing $n$ "successes" in $n$ trials, what is the probability of a success on the next try? Jeffreys expressed his view that his non-informative prior and Haldane's prior that are symmetric around a half and go to infinity at $p = 0$ and $p = 1$ put too much mass in the neighborhoods of the extreme points, 0 and 1, while the Laplace uniform prior does not put enough mass in the neighborhood of the extreme points, again very pragmatically suggests a modified Laplace mixed prior density. As I have pointed out in Zellner [(1997b), page 117], my "maximal data information prior" (MDIP) for the binomial parameter $p$, in the closed interval 0 to 1 is

$$f(p) = 1.6186 x p^p x (1-p)^{(1-p)}$$

a density that is symmetric about $p = \frac{1}{2}$, its minimal value, and rises to 1.6186 at both $p = 0$ and $p = 1$. It is thus "between" the uniform and Jeffreys's and Haldane's priors that shoot off to infinity at the end points. Further, a similar result is available for the multinomial model's parameters. Also, the criterion functional that is optimized to produce this MDIP for the binomial parameter and many others is an information criterion functional (see Zellner, 1997b, page 128ff for details) and uses of it to produce priors for many models and problems that are in general invariant to linear transformations and can be made invariant to other relevant transformations and related to work by Jeffreys, Berger, Bernardo and others on this difficult problem.

Further, in the case of a prior for a correlation coefficient in a normal model, the authors present an "arc-sine" prior for the correlation coefficient that is exactly the MDIP for this parameter. Also, the MDIP approach has been applied to the AR(1) stationary process, a problem discussed in the current paper and by Jeffreys. As explained in Zellner [(1997b), page 138], the MDIP for this problem is $p(b, \sigma) = c(1-b^2)^{1/2}/\sigma$, with $-1 < b < 1$. This contrasts markedly with the Jeffreys prior $p(b, \sigma) = c^1/(1-b^2)^{1/2}\sigma$ that the authors present without noting that Jeffreys [(1967), page 359] states, "The [Jeffreys] estimate rule gives a singularity at not only $b = 1$, which might be tolerable, but also at $b = -1$, which is not." Thus Jeffreys, always honest and pragmatic, reports that his prior for this problem is intolerable. See also the MDIP prior for parameters of a stationary AR(2) process and many other models in Zellner (1997a). Given the remarkable properties of MDIPs and the general principle from which they are derived, it is indeed surprising that the authors make no mention of them.

In closing, I shall quote the conclusions regarding Jeffreys's research contributions made by a leading Bayesian statistician to provide readers with an alternative appreciation of Jeffreys's contributions that can be compared to that presented in the authors' paper. Seymour Geisser (1980) wrote:

> If one were to present a short selected summary of Jeffreys's contributions to Bayesian inference, I believe that the following would be on everybody's list.
>
> (1) He made the inductive argument a "logical" one within the context of a Bayesian framework and maintained it could only be so within this framework.
> (2) He made a valiant attempt to quantify lack of knowledge by giving rather clever canonical rules and conventions but was not constrained to think only in these terms.
> (3) He produced a normative catalog of cogently reasoned Bayesian solutions to many conventional statistical paradigms.
> (4) He introduced and developed invariance considerations into the Bayesian system.
> (5) His devastating critiques of the various frequency theories propounded by Venn, Fisher, Neymann and others were, in the words of de Finetti (1970), closely argued and unanswerable.
>
> In summary, Jeffreys's approach amalgamated a Bayesian system with two primitive data principles reflective of public scientific work: (1) letting the data speak for themselves and (2) the actual units in which you choose to express your work should by and large not affect the inference. This is translated into so-called noninformative priors and invariance under suitable transformations. It was a rather remarkable conception, brilliantly executed, whose ultimate test is how it works in practice (19–20).

Thanks again to the authors for their many insightful comments that are very relevant for appraising Jeffreys's technical work and its mathematical basis. In this connection, some years ago I asked the famous statistician David Cox why the British have been so successful in the field of Statistics. He replied that British statisticians were well trained in applied mathematics, not theoretical mathematics. Perhaps this explains Jeffreys's limited knowledge of past measure theory and ignorance of recent results on alternative limiting processes for defining unbounded measures that have appeared since his death in 1989.